\documentclass[10pt]{article}

\usepackage{cite} 
\usepackage{url}  
\usepackage{ifthen}  
\usepackage{multicol}   
\usepackage[utf8]{inputenc} 
\usepackage{multirow}
\usepackage{epsfig}
\usepackage{subfigure}
\usepackage{verbatim}
\usepackage{rotating}

\usepackage[labelfont=bf,labelsep=period,justification=raggedright]{caption}

\makeatletter
\renewcommand{\@biblabel}[1]{\quad#1.}
\makeatother

\date{}

\begin{document}

\begin{flushleft}
{\Large
\textbf{Enhancing the functional content of protein interaction networks}
}
\\
Gaurav Pandey$^{1\ast}$, 
Sahil Manocha$^{2\dagger}$, 
Gowtham Atluri$^{3}$,
Vipin Kumar$^{3}$
\\
\bf{1} Institute for Genomics and Multiscale Biology and Department of Genetics and Genomic Sciences, Mount Sinai School of Medicine, New York, NY, USA
\\
\bf{2} Department of Computer Science and Engineering, Indian Institute of Technology, Kharagpur, India
\\
\bf{3} Department of Computer Science and Engineering, University of Minnesota, Minneapolis, MN, USA
\\
$\ast$ E-mail: gaurav.pandey@mssm.edu\\
$\dagger$ Affiliation while this work was done.
\end{flushleft}

\section*{Abstract}
Protein interaction networks are a promising type of data for studying complex biological systems. However, despite the rich information embedded
in these networks, they face important data quality challenges of noise and incompleteness that adversely affect the results obtained from their analysis. Here, we explore the use of the
concept of common neighborhood similarity (CNS), which is a form of local structure in networks, to address these issues. Although several CNS measures have been proposed 
in the literature, an understanding of their relative efficacies for the analysis of interaction networks has been lacking. 
We follow the framework of graph transformation to convert the given interaction network into a transformed network
corresponding to a variety of CNS measures evaluated. The effectiveness of each measure is then estimated by comparing the quality of protein function predictions obtained from its corresponding
transformed network with those from the original network. Using a large set of \emph{S. cerevisiae} interactions, and a set of $136$ GO terms,
we find that several of the transformed networks produce more accurate predictions than those obtained from the original network.
In particular, the $HC.cont$ measure proposed here performs particularly well for this task. Further investigation reveals that the two major factors
contributing to this improvement are the abilities of CNS measures, especially $HC.cont$, to prune out noisy edges and introduce new links between functionally related proteins.

\section*{Introduction}
Protein interaction networks are one of the most promising types of data for studying complex biological systems, as well as for addressing specific
problems, such as identifying disease-related proteins~\cite{chuang07} and finding functional modules and functions of individual proteins~\cite{pandey06survey,sharan07}.
In particular, since functionally related proteins tend to be highly inter-connected in these networks, several approaches, such as neighborhood-based prediction~\cite{schwik00}
and FunctionalFlow~\cite{nabieva05}, have been proposed for predicting the functions of unannotated proteins using this type of data. 

However, despite the rich information embedded in protein interaction networks, they face several data quality challenges that adversely
affect the results obtained from their analysis. One of the most prominent of these problems is that of noise in the data,
which manifests itself primarily in the form of spurious or false positive interactions~\cite{mering02,hart06}.
Studies have shown that the presence of noise in these networks has significant adverse affects on the performance of protein function prediction algorithms
\cite{deng03b}. Another important problem facing the use of these networks is their incompleteness, i.e., the absence of biologically valid interactions from the current interaction data sets
\cite{mering02,hart06,silva06}. This lack of completeness is mainly caused by the specific targeting of bait and prey proteins by individual studies (based on criteria such as functional
annotations), which, by its very nature, can only generate small samples of the entire interactome of an organism. Not surprisingly, the incompleteness of such
valuable data leads to missed biological insights that could have otherwise been gained if this data was available. Thus, although the numbers presented above are only estimates, it is clear that
noise (false positives) and incompleteness (false negatives) are major challenges facing protein interaction data that need to  be dealt with in order to obtain richer information from them.

Here, we study a set of techniques that make use of the (local) structure of an interaction network to address these issues. For the purpose of explaining and implementing these techniques, we represent a protein
interaction network as an undirected graph, with proteins being represented by nodes and interactions by edges\footnote{For this reason, the sets of terms ("network","graph"), 
("protein","node") and ("interaction","edge") will be used interchangeably in this paper.}. We also assume that weights reflecting the reliabilities of individual interactions
are assigned to the corresponding edges. Most traditional approaches for the analysis of protein interaction network are based on this representation, and focus on the direct interactions (edges
directly connecting two nodes) to conduct their analysis.

However, in addition to the direct interactions, the structure of the entire protein interaction network provides information about several other types of higher-level associations between proteins.
One of the most widely studied of these associations is that based on the idea of \emph{common neighborhood}~\cite{brun03,sam03,zhang05,chua06,pandey07}, where it is hypothesized that two
proteins that have several common direct neighbors (interaction partners) are likely to have a \emph{functional association} between them.
Consequently, several measures for the \emph{common neighborhood similarity} (CNS) of two proteins, based on different variants of the number of their common neighbors,
have been proposed. Several of these similarity measures have been used for clustering the proteins in the given network into functional modules~\cite{brun03,sam03,zhang05}, and many of the
resultant modules were determined to be hard to discover directly from the original network. Chua \emph{et al.}~\cite{chua06} used one such CNS measure, named FS (Functional Similarity), to predict the functions
of unannotated proteins, and their approach showed better performance than several other function prediction approaches. Pandey \emph{et al.}~\cite{pandey07} utilized some CNS measures within a graph transformation
procedure in the context of handling the noise and incompleteness issues with protein interaction data discussed above. The hypothesis underlying this work was that true interactions are
more likely between proteins that have a robust common neighbor configuration, and the interactions between proteins that do not participate in such a configuration are likely to be spurious.
Using the accuracy of protein function prediction as an evaluation criterion of the benefits of this CNS-based transformation, it was shown that more accurate predictions of protein function
could be obtained from many of the transformed networks as compared to the original one. In particular, the $h-confidence$ measure~\cite{xiong06b} produced the best performance
among all the CNS measures considered.

Despite the demonstration of the utility of the different CNS measures in various contexts, an understanding of their relative efficacies for the analysis of protein interaction networks has been
lacking due to several reasons. Firstly, as discussed above, each of these measures has been used for very different applications, that too on different interaction data sets, thus making their relative
comparison difficult. Furthermore, even in cases where these measures have been used in the context of function prediction~\cite{chua06,pandey07} or functional module discovery
\cite{sam03,zhang05}, different sets of functional classes and evaluation measures are used, making this comparison even harder. In this paper, we attempt to fill in this gap by conducting
an extensive comparative evaluation study of the CNS measures within the uniform context of protein function prediction from both unweighted and weighted interaction networks. We follow the systematic
framework of graph transformation~\cite{pandey07} to generate a transformed network corresponding to each of the CNS measures evaluated. The effectiveness of each measure is then estimated by comparing the quality of
function predictions made from their corresponding transformed network with those from the original network.

Using a large set of \emph{S. cerevisiae} interactions from the BioGRID database~\cite{stark2006biogrid}, and a set of $136$ GO Biological Process terms~\cite{myers06}, we find that
several of the transformed networks produce more accurate predictions than those obtained from the original network, although some networks based on binary CNS measures do not perform as well.
In particular, the $HC.cont$ measure proposed here performs particularly well from this perspective. An important contribution of our work is the
explanation of this variation in performance in terms of the different types of changes introduced into the network structure due to the transformation using the different measures.
This investigation yields that the ability of the CNS measures to identify and drop noisy edges is an important reason for the better predictions obtained after the transformation.
Further examination reveals that CNS measures are effective at introducing novel and accurate functional associations between proteins belonging to the same functional classes,
which in turn factors into the corresponding transformed networks performing better for function prediction than the original network. Interestingly,
the order of the performance of the CNS measures in these experiments matches that of their performance in function prediction experiments, with $HC.cont$ performing the best among all the
measures. Overall, these results are expected to provide a better understanding of the efficacy of CNS measures for handling data quality issues with protein interaction data and the
utility of these measures for enhancing the functional content of protein interaction data. 

Finally, before discussing our methods and results in detail, we would like to note that several other methods have also been proposed for assessing the reliabilities of protein interactions using other
data sources, such as microarray data and amino acid sequences~\cite{deane02,deng03b,suthram06}. However, since our focus is on using the information in the given interaction network itself for this
task, we do not evaluate these methods in this study. These two types of approaches provide complementary information about the reliability of an interaction,
and thus, their combination is expected to provide an even more accurate estimation of these reliabilities. However, this investigation is outside the scope of this paper.

\section*{Materials and Methods}
In this section, we will discuss the interaction data set, functional annotations, CNS measures and evaluation protocol used in this study.
\subsection*{Interaction data and functional annotations}
\label{sec:data}
We obtained our interaction data set from the BioGRID database~\cite{stark2006biogrid} in February, 2008. This data set included $34,483$ interactions between
$3774$ \emph{S. cerevisiae} proteins. In addition to using the unweighted (binary) version of this network, we also generated a weighted version, where each edge was assigned a weight
equal to the fraction of the total number of studies included in the data set ($26$) where it was detected. We also performed similar experiments on Collins \emph{et al.}'s high-confidence
protein interaction data set~\cite{collins07}.

The functional annotations for these proteins were taken from the GO database~\cite{ash00} in February, 2008. In particular, we used $136$ GO Biological Process
terms that Myers \emph{et al.}~\cite{myers06} had determined to be relevant for functional analyses of \emph{S. cerevisiae} data (at least $4$ votes) and had at least $10$ member proteins included
in our interaction data set. The sizes of these classes varied from $10-372$. 
\subsection*{Common Neighborhood Similarity (CNS) measures}
\label{sec:measures}
We evaluated a variety of CNS measures in our study, which are discussed in this section. For the purpose of defining each of these measures, we will use the following standard notation:

\begin{itemize}
\item $u$ and $v$ are the nodes between which the similarity is being computed.
\item $N_u$ and $N_v$ are the direct interaction partners of $u$ and $v$ respectively, and $N_{uv}=N_u\cap N_v$.
\item $a_{u,v}$ denotes the (positive) weight of the edge between $u$ and $v$.
\end{itemize}
We now define and discuss the CNS measures studied in detail.

\subsubsection*{Jaccard similarity}
One of the most commonly used measures for the similarity of two sets, $N_u$ and $N_v$ here, is the Jaccard coefficient~\cite{sardiu08}, which is defined as follows:
\begin{equation}
Jaccard(u, v) = \frac{|N_{uv}|}{|N_u \cup N_v|}
\end{equation}
The Jaccard coefficient measures how similar the two sets are, and assumes a value of $1$ only if $N_u=N_v$. However, in this form, it can only be used for unweighted graphs. Also, this
measure does not incorporate the presence or absence of an interaction between $u$ and $v$ ($a_{u,v}$) itself.

\subsubsection*{Pvalue}
Samanta \emph{et al.}~\cite{sam03} proposed a probabilistic measure for the statistical significance of the common neighborhood configuration of two nodes $u$ and $v$ in an unweighted graph. The value of
this measure, named $Pvalue$ here, is the $-log_{10}$ value of the probability of $u$ and $v$ having a certain number of common neighbors by random chance, and is defined as: 
\begin{equation}
Pvalue(u,v) = -log_{10}(p(N,|N_u|,|N_v|,|N_{uv}|))
\end{equation}
Here, $N$ is the total number of proteins in the network, and $p(N,|N_u|,|N_v|,|N_{uv}|)$ is computed on the basis of a Binomial distribution as: 
\begin{equation}
p(N,|N_u|,|N_v|,|N_{uv}|) = \frac{\left(\begin{array}{c}N \\|N_{uv}|\end{array}\right) \left (\begin{array}{c} N -|N_{uv}|\\|N_u|-|N_{uv}| \end{array} \right )\left( \begin{array}{c} N -|N_u|\\|N_v|-|N_{uv}|\end{array}\right)}{\left(\begin{array}{c} N \\|N_u|\end{array} \right)\left(\begin{array}{c} N \\|N_v|\end{array} \right)}
\end{equation}

Thus, $Pvalue$ is expected to have a high value (low value of $p$) for the non-random common neighbor configurations in a network. However, similar to $Jaccard$, this measure is unable to take edge
weights into account, thus losing information about the reliabilities of interactions over which the measure is computed. Another potential weakness of this measure is that it does not incorporate the
value of $a_{u,v}$. However, perhaps an even more important question is whether a measure of statistical significance, such as $Pvalue$, can be used as a measure of the strength
of the association between two proteins? Results presented in the subsequent sections attempts to answer this question.

\subsubsection*{Functional Similarity (FS)}
Chua \emph{et al.}~\cite{chua06} proposed a measure named Functional Similarity (FS) for measuring the common neighborhood similarity of two proteins in an interaction network. For an unweighted network ($0/1$ weights), this measure, referred to as $FS.binary$, can be defined as:
\begin{equation}
FS.binary(u, v) = \frac{2|N_{uv}|}{|N_u - N_v|+ 2|N_{uv}|+ \lambda _{u, v}}\times \frac{2|N_{uv}|}{|N_v - N_u|+ 2|N_{uv}|+ \lambda _{v, u}} 
\end{equation}
where $\lambda _{u,v} = max(0,n_{avg} - (|N_u - N_v|+|N_{uv}|))$ and $n_{avg}$ is the average number of neighbors of each protein in the network. The purpose of the $\lambda$ factor is to penalize the score between proteins pairs where at least one of the proteins has too few neighbors, since the score may not be very reliable in such a case. Note that unlike the other measures, the computation of $FS$ assumes that a protein, say $u$, is included in its direct neighborhood, i.e., $N_u$.

Essentially, $FS$ separates the (functional) similarity of two proteins into two probabilities that denote the conditional probabilities of $u$ and $v$ being functionally related given the neighborhoods of $u$ and $v$ respectively. Each of these conditional probabilities are computed as how similar the set of common neighbors of $u$ and $v$ ($N_{uv}$) is to the set of individual neighbors of $u$ ($N_u$) and $v$ ($N_v$). The final $FS$ score is obtained as a product of these probabilities, assuming that they are independent. 

Also, by using $\Sigma_{w \in N_u}a_{u,w}$ as the generalization of $N_u$ (similarly for $N_v$), and $\Sigma_{w \in N_{uv}}a_{u,w}a_{v,w}$ as the generalization for $N_{uv}$, a version of the $FS$ measure, named $FS.cont$, can be defined for a weighted interaction networks as follows:
\begin{equation}
FS.cont(u, v) = \frac{2 \Sigma _{w \in N_{uv}}a_{u,w}a_{v,w}}{\Sigma _{w \in N_u}a_{u,w} + \Sigma _{w \in N_{uv}}a_{u,w}a_{v,w}+ \lambda _{u, v}}\times \frac{2 \Sigma _{w \in N_{uv}}a_{u,w}a_{v,w}}{(\Sigma _{w \in N_v}a_{v,w} + \Sigma _{w \in N_{uv}}a_{u,w}a_{v,w}+ \lambda _{v, u}}
\end{equation}
Note that we used a similar definition of $\lambda_{v, u}$ as for the unweighted network case, while using the weighted versions of $n_{avg}$, $|N_u|$, $|N_v|$ and $|N_{uv}|$. Note that Chua \emph{et al.}~\cite{chua06} proposed a slightly different definition for $\lambda_{v, u}$ that assumes the knowledge of the functions of the proteins, which was not applicable in our case.

\subsubsection*{Topological Overlap Measure (TOM)}
This measure was proposed for network analysis by Ravasz \emph{et al.}~\cite{ravasz02} and was subsequently used for co-expression network analysis by Zhang and Horvath~\cite{zhang05}. TOM measures the strength of the association between two nodes in a graph based on the similarity of their common neighborhood to the smaller of the individual neighborhoods of the two nodes. For the case of an unweighted or binary network, the $TOM.binary$ measure can be defined as:
\begin{equation}
TOM.binary(u, v) = \frac{|N_{uv}|+ a_{u,v}}{min\{|N_u|,|N_v|\}+1 - a_{u,v}}
\end{equation}
It can be seen that the basic definition of $TOM.binary$ is quite straightforward. However, an important factor included in this measure is the presence or absence of an edge between $u$ and $v$ ($a_{u,v}=1$ and $0$ respectively) in the original network through the terms $a_{u,v}$ and $1-a_{u,v}$ in the numerator and denominator respectively. The inclusion of these factors has the desirable effect that the value of $TOM.binary$ is increased if $u$ and $v$ are known to have an interaction, which is sensible since the knowledge of this interaction should contribute favorably to the score for these proteins.

Again, using the same generalizations as for $FS.cont$ produces a formulation of $TOM$ for weighted networks, i.e. $TOM.cont$, as:
\begin{equation}
TOM.cont(u, v) = \frac{\Sigma _{w \in N_{uv}}a_{u,w}a _{v,w} + a _{u,v}}{min \{\Sigma _{w \in N_u}a _{u,w}, \Sigma _{w \in N_v}a _{v, w}\} + 1 - a _{u,v}}
\end{equation}

Zhang and Horvath~\cite{zhang05} and others~\cite{carlson06,horvath06,ghazal06} have used this measure extensively for analyzing gene co-expression networks in several studies. We considered this measure for transforming protein interactions networks.

\subsubsection*{H-Confidence (HC)}
Pandey \emph{et al.}~\cite{pandey07} demonstrated an innovative application of Xiong \emph{et al.}'s $h-confidence$ measure ($HC$) measure~\cite{xiong06b}, originally designed for the analysis of binary data matrices, to the pre-processing of protein interaction networks, both weighted and unweighted. We modified the original definition of $HC$~\cite{pandey07} slightly to define the $HC.binary$ measure as:
\begin{equation}
HC.binary(u, v) = \frac{|N_{uv}|+ a_{u,v}}{\max\{|N_u|,|N_v|\}}
\end{equation}

The change here is the addition of the $a_{u,v}$ term in the numerator to incorporate the presence/absence of the interaction between $u$ and $v$. As per this definition, $HC.binary$ rewards cases where the set of common neighbors ($N_{uv}$) is very similar to the sets of individual neighbors of $u$ and $v$. However, due to the use of the $\max\{|N_u|,|N_v|\}$ term in the numerator, $HC.binary$ penalizes the cases where the degree of at least one of the nodes is substantially higher than $|N_{uv}|$, thus avoiding a bias in favor of high-degree or hub nodes in the network. This behavior of $HC.binary$ is in sharp contrast to that of the similarly defined $TOM.binary$ measure, the value of whose denominator is generally small for protein interaction networks due to the use of the $\min\{|N_u|,|N_v|\}$ term and the fact that a vast majority of the nodes in these networks have very small degrees.

Finally, using the same generalizations as for $FS$ and $TOM$, the definition of $HC.binary$ can be extended to $HC.cont$ for the case of weighted interaction networks as follows:
\begin{equation}
HC.cont(u, v) = \frac{\Sigma _{w \in N_{uv}}a_{u, w}a_{v,w} + a_{u,v}}{max\{\Sigma _{w \in N_u}a_{u,w}, \Sigma _{w \in N_v}a _{v,w}\}}
\label{eqn:hccont}
\end{equation}
This definition of $HC.cont$ enables a more conservative estimation of $HC$-based common neighborhood similarity due to the use of the sum of the product of the edge weights, both of which are at most $1$ and thus their product is expected to be much smaller than the minimum of the two values. It should be noted that $HC.cont$ also has a behavior similar to $HC.binary$, wherein nodes with low weighted degrees in the original network are more likely to have links with higher $HC.cont$ scores as compared to higher weighted degree nodes in the original network.

As can be seen, these measures adopt different formulations for computing common neighborhood similarity between two nodes (proteins) in a graph (interaction network). We next describe how we evaluated these measures within the frameworks of graph transformation and protein function prediction.

\subsection*{Evaluation methodology}
\label{sec:evalmethod}
Our evaluation methodology consists of the following two steps:
\begin{itemize}
\item First, each of the above CNS measures is used to compute the similarity (strength of the association) between each pair of proteins in the input interaction network, depending on whether
they operate on the weighted
or unweighted version of the network. Next, a threshold is chosen for each score such that the number of pairs with a score higher than this threshold is as close as possible to the number of interactions
in the original network. The pairs that score higher than the threshold are structured as a network, and constitute the \emph{transformed network} for the corresponding measure. Note that this
form of thresholding helps us reduce the bias in the performance of the function prediction algorithms (described next) due to the size of the network they are run on.
\item Next, two different protein function prediction algorithms are run on the original as well as the transformed networks to make predictions over a set of $136$ GO BP process terms/classes.
The first algorithm used was Nabieva \emph{et al.}'s FunctionalFlow algorithm~\cite{nabieva05}. We also used a simple neighborhood-based algorithm
inspired by Schwikowski \emph{et al.}'s function prediction algorithm~\cite{schwik00}. Here, the likelihood score of a query protein performing certain function is simply counted as the sum of the weights of its
interactions with proteins that are known to be annotated with that function, and these scores are collected for all the unannotated proteins in the data set for all the relevant functions.
The predictions from both these algorithms are evaluated within a five-fold cross-validation setup by computing the Area Under the ROC Curve (AUC) score for each class separately.
\end{itemize}
The results obtained from this methodology are discussed in the next section.

\section*{Results}
In this section, we will discuss the results of our evaluation study, and also the subsequent analyses that we carried out to explain the observed trends.

\begin{table}[t!]
\centering
\begin{tabular}{|c|c|c|c|}\hline
\multirow{2}{*}{{\bf CNS Measure}} & \multirow{2}{*}{\# {\bf Interactions}} & \# {\bf Connected} & {\bf Range of}\\
& & {\bf proteins} & {\bf edge weights}\\\hline
$Jaccard$ & $34637$ & $3492$ & $0.13-1$\\\hline
$Pvalue$ & $34481$ & $2162$ & $6.61-317.13$\\\hline
$FS.binary$ & $34481$ & $2359$ & $0.08-0.93$\\\hline
$TOM.binary$ & $34483$ & $3646$ & $0.5-1$\\\hline
$HC.binary$ & $34496$ & $3559$ & $0.19-1$\\\hline
$FS.cont$ & $34483$ & $3745$ & $0.0011-0.41$\\\hline
$TOM.cont$ & $34474$ & $3774$ & $0.0173-0.46$\\\hline
$HC.cont$ & $34487$ & $3757$ & $0.0135-1$\\\hline
\end{tabular}
\caption{Details of transformed networks produced using different CNS measures.}
\label{table:trans_network_details}
\end{table}

\subsection*{Details of transformed networks}
Table~\ref{table:trans_network_details} lists the details of the different transformed networks generated using the methodology described above.
As can be seen, the number of interactions in these networks, as well as the number of connected proteins with at least one interaction, are almost the same as the original
network, thus ensuring that the downstream analysis of these networks is not biased due to a variation in these factors. The only exceptions to these observations are the $Pvalue$
and $FS.binary$ networks, where the number of connected proteins is substantially lower than the original network. This discrepancy occurs primarily due to low scores being assigned to edges involving
the weakly connected nodes in the original network according to these measures, which are not included in the transformed networks obtained after thresholding the full set of scores.
Also, since there are a large number ($30001$) of protein pairs with a $TOM.binary$ score of $0.5$, we had to randomly choose $21210$
pairs out of this set to create a network with the same number of interactions as the original one.

\subsection*{Performance of function prediction algorithms}
\label{sec:function_prediction}
We evaluated the utility of each of the transformed networks for predicting the membership of \emph{S. cerevisae} proteins in $136$ GO Biological Process classes, and compared their
performance with the weighted ($Original.cont$) and unweighted ($Original.binary$) versions of the original interaction network. Tables~\ref{table:BiogridFF} and~\ref{table:BiogridNbd}
detail the results of this evaluation using the FunctionalFlow and neighborhood-based function prediction methods respectively. The following consistent observations can be made from these tables:
\begin{itemize}
\item Suprisingly, the $Pvalue$ and $Jaccard$ measures, which have been previously used~\cite{sam03,sardiu08} for the analysis of unweighted interaction networks (similar to our
$Original.binary$ network), produce substantially worse predictions than the $Original.binary$ network itself. This is primarily due to their inability to incorporate real-valued
edge weights, as well as the weight of the edge between the pair of proteins being evaluated. In particular, the performance of $Pvalue$ is likely to be adversely affected by the wide scale of
scores assigned to the edges in its transformed network (Table~\ref{table:trans_network_details}), and also indicates the limitations of using a measure of statistical significance to estimate
the strength of an association between proteins.
\item For the other measures ($FS$, $TOM$ and $HC$), the transformed network generated from $Original.cont$ produce much better results than the ones generated
from the $Original.binary$ network, since the latter are unable to utilize the edge reliability scores. In fact, this observation
is also true for the original network, which is the reason we choose the $Original.cont$ network as the benchmark for comparing the performance of the CNS measures.
\item Among all the measures, it can be seen that only the measures that can utilize edge weights, namely $FS.cont$, $TOM.cont$ and $HC.cont$, perform better than (or almost the same as) the
$Original.cont$ network in terms of the mean AUC score across all the classes. Overall, this result shows that it is possible to perform more
accurate analysis on the original interation data by transforming it using appropriate CNS measures.
\item Among the continuous CNS measures, $HC.cont$ performs the best in terms of almost all the evaluation metrics. In addition to producing the highest mean AUC increase, $HC.cont$ is also able
to substantially increase the AUC (increase$>$0.05) for a much larger number of classes, as compared to those for which it leads to a major decrease in AUC (decrease$>$0.05). 
This performance is because $HC.cont$ is better able to synthesize the common neighborhood configuration of two proteins, i.e., the connecting edges and their weights,
into an accurate measure for the similarity or the strength of association of the two proteins.
\end{itemize}

\begin{table*}[t!]
\centering
{\footnotesize \begin{tabular}{|c|c|c|c|c|c|c|c|}\hline
{\bf CNS Measure} & {\bf Mean} & {\bf Mean AUC} & {\bf Max AUC} & {\bf \# Classes} & {\bf \# Classes} & {\bf Max AUC} & {\bf \# Classes}\\
& {\bf AUC} & {\bf Change} & {\bf Increase} & {\bf Increase} & {\bf with AUC} & {\bf Decrease} & {\bf with AUC}\\
& & & & & {\bf increase$>$0.05} & & {\bf decrease$>$0.05}\\\hline
$Original.cont$ & $0.7944$ & & & & & &  \\\hline
$FS.cont$ & $0.7941$ & $-0.0003$ & $0.0846$ & $83$ & $3$ & $0.0662$ & $5$ \\\hline
$TOM.cont$ & $0.8034$ & $0.0089$ & $0.0636$ & $105$ & $2$ & $0.1782$ & $3$ \\\hline
$HC.cont$ & $0.8266$ & $0.0281$ & $0.2244$ & $109$ & $34$ & $0.1199$ & $6$ \\\hline
$Original.binary$ & $0.7847$ & $-0.0097$ & $0.0502$ & $48$ & $1$ & $0.2222$ & $8$ \\\hline
$Jaccard$ & $0.7291$ & $-0.0653$ & $0.2738$ & $28$ & $18$ & $0.4759$ & $86$\\\hline
$Pvalue$ & $0.6374$ & $-0.1570$ & $0.1740$ & $5$ & $3$ & $0.5691$  & $124$\\\hline
$FS.binary$ & $0.6750$ & $-0.1195$ & $0.1740$ & $5$ & $1$ & $0.4945$ & $113$ \\\hline
$TOM.binary$ & $0.7365$ & $-0.0579$ & $0.3332$ & $23$ & $9$ & $0.3227$ & $78$ \\\hline
$HC.binary$ & $0.7578$ & $-0.0367$ & $0.2989$ & $35$ & $25$ & $0.3773$ & $70$ \\\hline
\end{tabular}}
\caption{Performance statistics of FunctionalFlow over the original and several transformed interaction networks. All the increase/decrease results are with respect to the $Original.cont$ network.}
\label{table:BiogridFF}
\end{table*}

\begin{table*}[t!]
\centering
{\footnotesize \begin{tabular}{|c|c|c|c|c|c|c|c|}\hline
{\bf CNS Measure} & {\bf Mean} & {\bf Mean AUC} & {\bf Max AUC} & {\bf \# Classes} & {\bf \# Classes} & {\bf Max AUC} & {\bf \# Classes}\\
& {\bf AUC} & {\bf Change} & {\bf Increase} & {\bf Increase} & {\bf with AUC} & {\bf Decrease} & {\bf with AUC}\\
& & & & & {\bf increase$>$0.05} & & {\bf decrease$>$0.05}\\\hline
$Original.cont$ & $0.8028$ & & & & & & \\\hline
$FS.cont$ & $0.8101$ & $0.0073$ & $ 0.0371 $ & $108$ & $0$ & $0.0368$ & $0$ \\\hline
$TOM.cont$ & $0.8095$ & $0.0067$ & $0.0522 $ & $103$ & $1$ & $0.0288$ & $0$ \\\hline
$HC.cont$ & $0.8189$ & $0.0161$ & $0.1423 $ & $99$ & $14$ & $0.0960$ & $4$ \\\hline
$Original.binary$ & $0.7973$ & $-0.0055$ & $0.0602$ & $28$ & $1$ & $0.0561$ & $1$ \\\hline
$Jaccard$ & $0.7296$ & $-0.0732$ & $0.1426$ & $15$ & $8$ & $0.2407$ & $95$ \\\hline
$Pvalue$ & $0.6863$ & $-0.1165$ & $0.1157$ & $5$ & $1$ & $0.3501$  & $121$\\\hline
$FS.binary$ & $0.7242$ & $-0.0786$ & $0.1028$ & $7$ & $1$ & $0.3147$ & $98$ \\\hline
$TOM.binary$ & $0.7162$ & $-0.0866$ & $0.1512$ & $9$ & $3$ & $0.3071$ & $108$ \\\hline
$HC.binary$ & $0.7535$ & $-0.0492$ & $0.1165$ & $23$ & $10$ & $0.2349$ & $69$ \\\hline
\end{tabular}}
\caption{Performance statistics of neighborhood-based function prediction over the original and several transformed interaction networks. All the increase/decrease results are with respect to the $Original.cont$ network.}
\label{table:BiogridNbd}
\end{table*}

We conducted similar experiments on Collins \emph{et al.}'s high-confidence protein interaction data set~\cite{collins07}, and the results are detailed in Section 1 of Supplementary Results.
Here, $HC.cont$ was the only measure able to produce more accurate predictions than the original high-confidence network, thus adding further credibility to its utility for enhancing the functional content
of protein interaction data.

Next, we conducted an extensive investigation to explain these variations in the performance of the continuous CNS measures, namely $FS.cont$, $TOM.cont$ and $HC.cont$, in terms of the changes they
introduce into the network structure. The following subsections detail the results of this investigation on the BioGrid data set.

\begin{figure}[t!]
\centering
\subfigure[Comparison of degree distributions for $FS.cont$.]{\epsfig{file=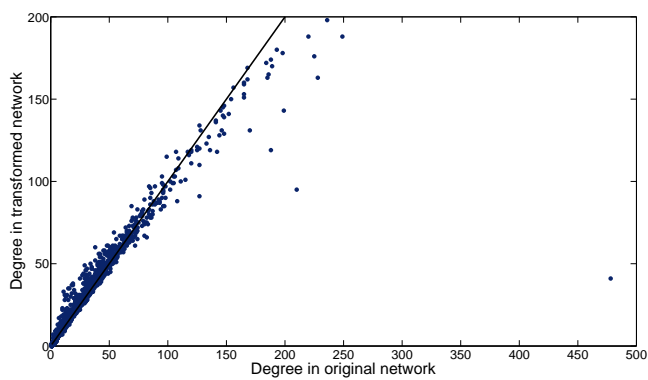,scale=0.2}\label{fig:fscont_degrees}}
\subfigure[Comparison of degree distributions for $TOM.cont$.]{\epsfig{file=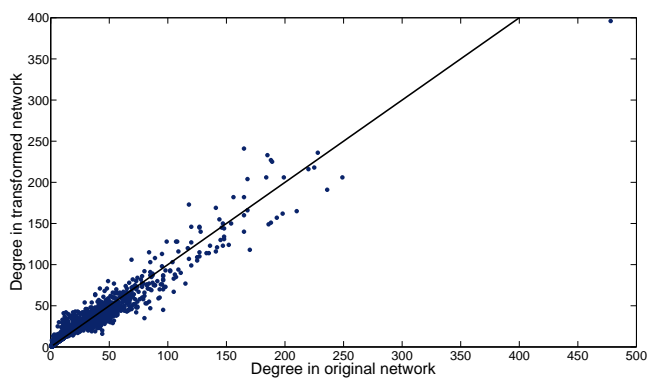,scale=0.2}\label{fig:tomcont_degrees}}
\subfigure[Comparison of degree distributions for $HC.cont$.]{\epsfig{file=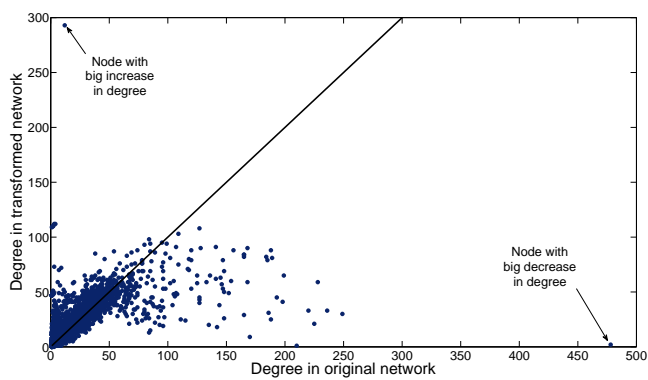,scale=0.2}\label{fig:hccont_degrees}}
\caption{Scatter plots comparing the degrees of the nodes in the orginal and the transformed network created using different continuous CNS measures (Plots best seen in color).}
\label{fig:degree_distribution}
\end{figure}

\subsection*{Changes in network structure}
\label{sec:network_changes}
It can be observed from the description of the CNS measures that two proteins that are not even connected in the original network may have a high CNS score, and vice versa,
depending on their common neighborhood configuration. The natural consequence of this is that the structure of the resultant transformed network(s) may be substantially different from that of the original
network. Indeed, the improvement in the function prediction results for some of the CNS measures can be attributed largely to these changes in the network structure. Thus, in this part of the study, we
focused on identifying the most prominent of these changes introduced by $TOM.cont$, $FS.cont$ and $HC.cont$.

To identify these changes, we compared the degree of each node in the original and transformed networks produced by these measures, and Figure \ref{fig:degree_distribution} shows this comparison
through scatter plots. As can be seen from the plots
in Figure~\ref{fig:fscont_degrees} and~\ref{fig:tomcont_degrees}, the degrees of the nodes remain largely unchanged (close to the $y=x$ line) between the original and
the transformed networks produced by $FS.cont$ and $TOM.cont$ respectively. This indicates the tendency of these measures to maintain the network structure and focus on
assigning more accurate reliability scores to the interactions that lead to an improvement in the results of protein function prediction.

In contrast, Figure~\ref{fig:hccont_degrees} shows that there is a substantial difference between the degrees of several nodes in the original and the $HC.cont$-transformed network.
These differences, which include examples of both decrease as well as increase in node degrees after the transformation, can be explained on the basis of the formula for $HC.cont$
(Equation~\ref{eqn:hccont}). Here, the denominator is the maximum of the (weighted) degrees of the two nodes, which implies that unless the numerator
has a high value, this measure will assign a low score to protein pairs where at least one of the proteins has a high (weighted) degree. This effect leads to the observed change in degrees.
For instance, consider the case of a node with degree $478$ in the original network (point at bottom right corner of Figure~\ref{fig:hccont_degrees}),
all of whose edges, except $13$, had low weights (less than $0.1$). The
result of this configuration is that the $HC.cont$ scores involving this node are low, since the numerator of Equation~\ref{eqn:hccont} is small because of the low edge weights, and the
denominator is high because of the high degree of this node. Indeed, after the $HC.cont$-based transformation, the highest weight of an edge involving this node in the transformed network is only
$0.0209$. As a result of this, all but $2$ of these intermediate edges are pruned out to retain the size of the original network. On the other hand, one of the nodes connected to this high-degree
node, which had $12$ neighbors in the original network (point at top left corner of Figure~\ref{fig:hccont_degrees})
obtained $293$ neighbors after the $HC.cont$-based transformation, due to its much lower degree (lower value of the denominator in Equation~\ref{eqn:hccont})
and the higher weights of its edges (higher value of the numerator in Equation~\ref{eqn:hccont}). Interestingly, $284$ of the new neighbors of this node were neighbors of the high-degree node in
the original network, and these new edges were formed due the the presence of the latter as one of the common neighbors. Such a change in
configuration at some places in the original network led to the observed changes in the degree distributions between the original and $HC.cont$-based transformed networks.

Now, a natural question to ask here is whether these major changes in the network structure, namely the dropping and introduction of edges, are responsible for the improvements observed in the
function prediction results? We attempt to answer this question by studying the two cases separately in the following subsections.

\begin{figure}[t!]
\centering
\epsfig{file=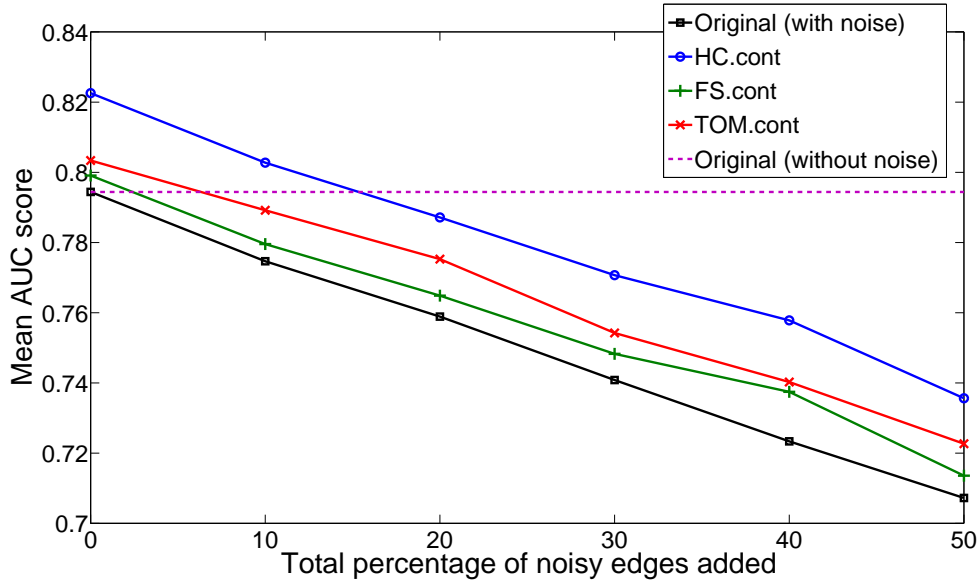,scale=0.3}
\caption{Performance of FunctionalFlow (in terms of the mean AUC score) on the original and resultant transformed networks at different levels of noise.}
\label{fig:auc_vs_noise}
\end{figure}

\subsection*{Robustness of CNS measures to noise}
\label{sec:noise_simulation}
One of the hypotheses underlying the use of common neighborhood similarity information is that it can be used for filtering out noisy or spurious interactions in a network, since two proteins connected
by a spurious interaction are less likely to have a larger number of common neighbors than two proteins connected by a true interaction. We explored this hypothesis as one of the benefits CNS measures
may provide for analyzing interaction networks. However, since it is difficult to identify the noisy edges in the original network apriori, we followed a simulation-based methodology for validating
this hypothesis. Under this methodology, we generated several randomly perturbed versions of the $Original.cont$ network, where the random rewiring model~\cite{guelzim2002topological,milo02}, where two
edges in the original network are chosen randomly and two new edges are created by swapping their end points. The weights of the original edges are also randomly
reassigned to the new edges. Applying this model to $10\%$, $20\%$, $30\%$, $40\%$ and $50\%$ of the edges in the original network gave us several "noisy" versions of the network, and we transformed
each of these networks using the $FS.cont$, $TOM.cont$ and $HC.cont$ measures.

In the first part of this analysis, we studied how the extent of noise in the noisy networks and their transformed versions affected the performance of the FunctionalFlow algorithm, measured in terms
of the average of the AUC scores of all the $136$ classes (GO terms). Figure~\ref{fig:auc_vs_noise} shows the results of this analysis as the noisy fraction of the network ranges from $10\%$ to $50\%$. As
expected, the results from all the networks become worse as the extent of noise increases. However, it is encouraging that all the transformed networks are able to resist the effect of noise to some
extent, and thus produce more accurate predictions than their corresponding noisy networks. $HC.cont$ (blue line) is consistently the best performer in this evaluation, and can produce a
performance as good as the original network (dotted purple line) even when almost $15\%$ of the edges in the network are spurious. The corresponding fractions for $TOM.cont$ and $FS.cont$ are only
about $5\%$ and $2\%$ respectively. Interestingly, the order of performance of these measures is the same as the function prediction results in Table~\ref{table:BiogridFF}, and close to that in Table
\ref{table:BiogridNbd}. Overall, these results demonstrate the robustness of the CNS-based transformed networks, particularly the one generated using
$HC.cont$, to the presence of noise in the original network, and serves as an important factor behind the improvement of function prediction results using these measures.

Furthermore, the setup of this simulation experiment allows us to examine in detail the precise changes being introduced into the network during the graph transformation process at different levels
of noise. For this purpose, we studied the extents of changes in terms of three different types of edges dropped from the original network and introduced into the transformed networks. From the results
of this analysis presented in Section 2 of Supplementary Results, it can be seen that several major changes are made to the original network during the graph transformation process by all the CNS measures.
It is particularly interesting to note that the ordering of these measures in terms of the extents to which they introduce these changes, namely $HC.cont$, $TOM.cont$ and $FS.cont$, is the same as that
for the function prediction results. This observation motivates the natural examination of how these changes in the network structure influence the
functional content of the original interaction network, the results of which are presented in the following subsection.

\subsection*{Enhancement of functional coherence}
In this part of the study, we investigated how these
changes in the network structure affect the functional content of the resultant network that leads to the observed improvements. To begin, we examined the functional coherence of the different
types of edges (Common, Dropped and Added) that are involved in these changes in the network structure. From the results of this analysis presented in Section 3 of Supplementary Results, it can be
seen that despite substantial variation in the fractions and functional coherence of the different sets of specific types of edges, the resultant transformed network produced by $HC.cont$ is the most
functionally coherent, followed by $TOM.cont$ and $FS.cont$.

\begin{figure}[t!]
\centering
\subfigure[Results using $HC.cont$.]{\epsfig{file=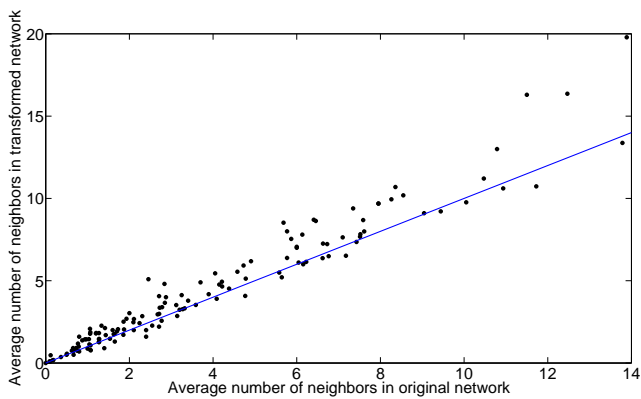,scale=0.2}}
\subfigure[Results using $TOM.cont$.]{\epsfig{file=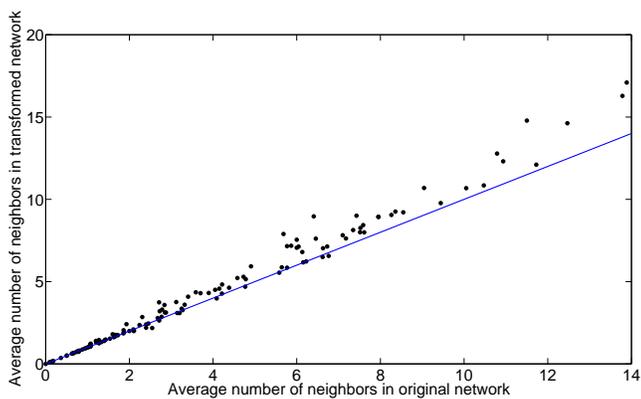,scale=0.2}}
\subfigure[Results using $FS.cont$.]{\epsfig{file=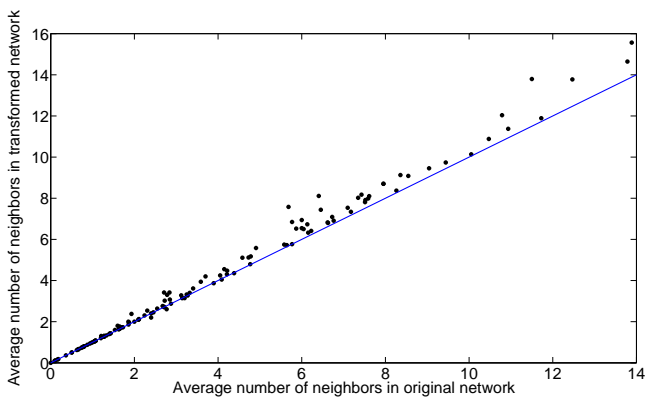,scale=0.2}}
\caption{Comparison of the functional connectivity between member proteins of individual functional classes between the original and the transformed networks generated using different CNS measures.}
\label{fig:neighbor_distribution}
\end{figure}

\begin{table}[t!]
\centering
\begin{tabular}{|c|c|c|}\hline
{\bf CNS measure} & {\bf Mean increase in} & {\bf Median increase in}\\
 & {\bf functional connectivity} & {\bf functional connectivity}\\\hline 
$HC.cont$ & $0.5489$ & $0.295$\\\hline
$TOM.cont$ & $0.4028$ & $0.102$\\\hline
$FS.cont$ & $0.2495$ & $0.0748$\\\hline
\end{tabular}
\caption{Summary statistics about change in functional connectivity after transformation using different CNS measures.}
\label{table:neighbor_distribution}
\end{table}

Given this global view of how the CNS measures enhance the overall functional relevance of the given interaction network, we next investigated how graph transformation using these measures affects
the functional connectivity between proteins belonging to individual classes using the following methodology. For each protein belonging to each of the $136$ classes, we computed the number
of their direct neighbors that belong to the same class in the original and the transformed networks, and used the average of this set of numbers as a measure of the functional connectivity of
each class in these networks. Figure~\ref{fig:neighbor_distribution} shows scatter plots comparing the values of these measures between the original and the transformed networks, and Table
\ref{table:neighbor_distribution} provides some summary statistics about these plots. While all the measures led to statistically significant ($p<10^{-10}$, Wilcoxon signed-rank test) increases
in the functional connectivity, it can be seen from these plots and the table that $HC.cont$ provides the highest overall per-class increase
in functional connectivity, followed by $TOM.cont$ and $FS.cont$. This order is identical to that of the performance of the function prediction algorithms on the corresponding transformed networks
(Tables~\ref{table:BiogridFF} and~\ref{table:BiogridNbd}, and thus provides another explanation for how $HC.cont$ helps improve the accuracy of function predictions by enhancing the connectivity
between proteins of the same functional class. In fact, this improvement is particularly high for the classes whose functional connectivity was very low in the original network.
For instance, for the $53$ classes whose connectivity was less than $2$ in the original network and for which there was a substantial increase in
functional connectivity due to $HC.cont$, the average improvement in AUC of FunctionalFlow predictions was $0.0388$ as against $0.0213$ for all the other classes. In comparison, transformation using
$FS.cont$ and $TOM.cont$ has very little effect on functional connectivity of these classes, and hence the accuracy of the predictions made (average change in AUC of FunctionalFlow predictions
was only $-0.0031$ and $0.0045$ respectively). This analysis shows that $HC.cont$ particularly helps improve the predictions for classes with poor functional connectivity for which it is difficult
to make accurate predictions from the original network.

In summary, the results in this section showed that an important factor behind the improved function prediction results obtained after graph transformation using the different continuous ($*.cont$)
CNS measures, particularly $HC.cont$, is the enhanced connectivity between functionally connected proteins.

Viewing these results in unison with those presented earlier shows that the CNS measures that utilize the real-valued edge reliability scores in the original network are effective
at identifying and dropping noisy edges, as well as introducing new edges between functionally associated proteins. It can be seen that these two operations are directly addressing the noise and
incompleteness issues with protein interaction data, which was the goal of this study. In particular, $HC.cont$ is able to address both these issues most effectively, and consequently leads to the
most accurate predictions of protein function. Similar advantages are provided by other CNS measures as well, namely $FS.cont$ and $TOM.cont$, although to smaller extents.  
\section*{Discussion}
In this paper, we evaluated the use of a variety of \emph{common neighborhood similarity} (CNS) measures to quantify the relationship of two proteins based on their common neighborhood, and
used them within the framework of graph transformation for the task of pre-processing protein interaction networks. 
We showed that such pre-processing, especially using CNS measures that take advantage of the real-valued edge reliability scores (weights), is able
to substantially improve the accuracy of predictions made for several GO Biological Process terms by standard protein function prediction algorithms. In particular, the continuous version of the
$h-confidence$ measure ($HC.cont$) produces the largest improvement in the prediction performance. We also investigated the structural changes introduced into the original network when it is
transformed using these CNS measures, especially $HC.cont$, in order to find the structural factors contributing to this improvement. We found that the two major factors contributing to this
improvement are abilities of $HC.cont$ and the other measures to prune out edges likely to be spurious (noisy) and introduce new links between functionally related proteins during the graph
transformation process. Overall, the methods and results of this study should help researchers adopt robust pre-processing schemes for protein interaction
networks, which should in turn help them obtain more accurate inferences from this type of data.

This work can be extended in several directions. Among the most direct extensions would be a validation of the noisy edges removed and the functional linkages added to the network during the graph
transformation process using experimental PPI assessment methods, such as that of \cite{braun08}.  Another direction would be to examine how the CNS measures evaluated here perform for other types of
network data, such as genetic interaction networks~\cite{tong04}, which have their own characteristics, such as the presence of both positively and negatively weighted edges. Finally, since
all the CNS measures considered here have different properties and different performance as a result, it is possible to develop hybrid CNS measures that combine the best properties of all these measures.

\section*{Acknowledgement}
We are thankful to Limsoon Wong for answering our queries about Functional Similarity ($FS$). This work was supported by NSF grant \# IIS-0916439, and a Doctoral Dissertation Fellowship from the University
of Minnesota Graduate School to GP.

\bibliographystyle{plos2009}
\bibliography{function,genex,genome,lit,multiple,phylo,pin,seq,str}

\newpage

\section*{Supplementary Results}

\subsection*{Function prediction experiments on Collins \emph{et al.}'s high-confidence interaction data set}

\begin{table*}[h!]
\centering
{\footnotesize \begin{tabular}{|c|c|c|c|c|c|c|c|}\hline
{\bf CNS Measure} & {\bf Mean} & {\bf Mean AUC} & {\bf Max AUC} & {\bf \# Classes} & {\bf \# Classes} & {\bf Max AUC} & {\bf \# Classes}\\
& {\bf AUC} & {\bf Change} & {\bf Increase} & {\bf Increase} & {\bf with AUC} & {\bf Decrease} & {\bf with AUC}\\
& & & & & {\bf increase$>$0.05} & & {\bf decrease$>$0.05}\\\hline
$Original.cont$	&0.7878	& & & & & & \\\hline					
$FS.cont$	&0.726	&-0.0618	&0.0807	&11	&3	&0.2357	&56 \\\hline
$TOM.cont$	&0.7664	&-0.0213	&0.116	&28	&8	&0.1663	&24 \\\hline
$HC.cont$	&0.8119	&0.0242	&0.1976&	68	&21	&0.0822	&6 \\ \hline
$Original.binary$	&0.7741&	-0.0137&	0.0929	&47&	5&	0.2482	&14 \\\hline
$Jaccard$	&0.7285	&-0.0592	&0.0869	&11	&3	&0.235	&49 \\\hline
$Pvalue$	&0.5597	&-0.2281	&0.0509	&2	&1	&0.4996	&93 \\\hline
$FS.binary$	&0.6204	&-0.1673	&0.0738	&3	&1	&0.4229	&88 \\\hline
$TOM.binary$	&0.7735	&-0.0143	&0.2327	&44	&6	&0.1913	&21 \\\hline
$HC.binary$	&0.7833	&-0.0044	&0.1311	&44	&14	&0.1357	&17 \\\hline
\end{tabular}}
\caption{Performance statistics of FunctionalFlow over the original and several transformed interaction networks. All the increase/decrease results are with respect to the $Original.cont$ network.}
\label{table:CollinsFF}
\end{table*}

\begin{table*}[h!]
\centering
{\footnotesize \begin{tabular}{|c|c|c|c|c|c|c|c|}\hline
{\bf CNS Measure} & {\bf Mean} & {\bf Mean AUC} & {\bf Max AUC} & {\bf \# Classes} & {\bf \# Classes} & {\bf Max AUC} & {\bf \# Classes}\\
& {\bf AUC} & {\bf Change} & {\bf Increase} & {\bf Increase} & {\bf with AUC} & {\bf Decrease} & {\bf with AUC}\\
& & & & & {\bf increase$>$0.05} & & {\bf decrease$>$0.05}\\\hline
$Original.cont$	&0.784	& & & & & &  \\\hline						
$FS.cont$	&0.7364	&-0.0476	&0.013	&5	&0	&0.1968	&37 \\\hline
$TOM.cont$	&0.7617	&-0.0223	&0.024	&16	&0	&0.1495	&12 \\\hline
$HC.cont$	&0.7932	&0.0092	&0.115	&60	&8	&0.0456	&0 \\ \hline
$Original.binary$	&0.7884	&0.0044	&0.1648	&60	&2	&0.0458	&0 \\\hline
$Jaccard$	&0.7244	&-0.0596	&0.0697	&10	&2	&0.2298	&44 \\\hline
$Pvalue$	&0.5765	&-0.2075	&-0.037	&0	&0	&0.4829	&97 \\\hline
$FS.binary$	&0.6396	&-0.1444	&-0.0285	&0	&0	&0.3161	&90 \\\hline
$TOM.binary$	&0.7713	&-0.0127	&0.1076	&27	&1	&0.1495	&10 \\\hline
$HC.binary$	&0.7808	&-0.0032	&0.1294	&43	&6	&0.0827	&7 \\\hline
\end{tabular}}
\caption{Performance statistics of neighborhood-based function prediction over the original and several transformed interaction networks. All the increase/decrease results are with respect to the $Original.cont$ network.}
\label{table:CollinsNbd}
\end{table*}

We applied the evaluation methodology described in the main text to Collins \emph{et al.}'s high-confidence data set, which consisted of $9064$ interactions covering $1620$ \emph{S. cerevisiae} proteins, and
evaluated the performance of FunctionalFlow and neighborhood-based function prediction algorithms over $98$ (of the original $136$) GO BP terms that had at least $10$ members each in this data set. Tables
\ref{table:CollinsFF} and~\ref{table:CollinsNbd} detial the results of these experiments in the same manner as Tables $2$ and $3$ in the main text. Interestingly, none of the measures other than $HC.cont$
were able to consistenly improve the overall AUC of the classes, especially because of the high quality of the network and thus the difficulty of improving the results over that obtained from the $Original.cont$
network. However, $HC.cont$ outperforms $Original.cont$ on all the metrics, thus demonstrating that it is capable of extracting rich functional information even from highly refined protein interaction networks.

\subsection*{Extents of changes to network structure}
\begin{figure}[h!]
\centering
\subfigure[Percentage of noisy edges removed during graph transformation at different levels of noise.]{\epsfig{file=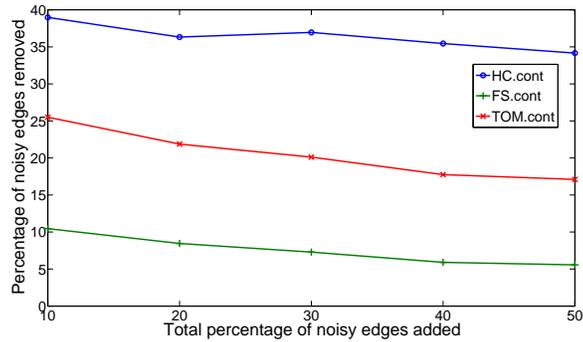,scale=0.18}\label{fig:removed_vs_noise}}
\subfigure[Percentage of edges in the original network that were dropped during graph transformation at different levels of noise.]{\epsfig{file=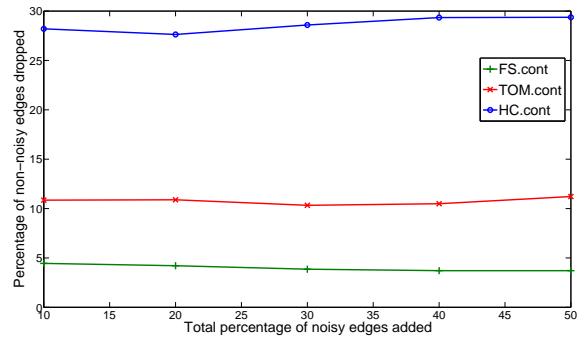,scale=0.18}\label{fig:valid_edges_dropped}}
\subfigure[Percentage of new edges introduced into the transformed networks at different levels of noise.]{\epsfig{file=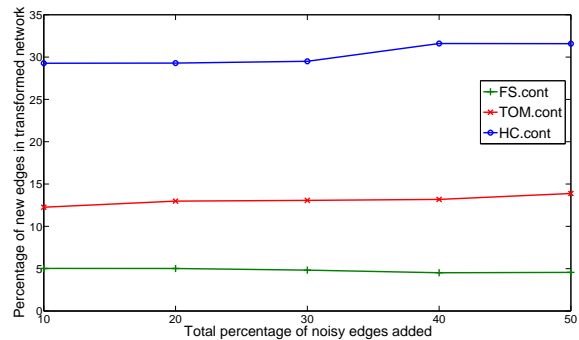,scale=0.18}\label{fig:fraction_edges_added}}
\caption{Analysis of the robustness of the original interaction network and its transformed versions in terms of the extents of the changes introduced into the network during transformation (Plots best viewed in color).}
\label{fig:noise_analysis}
\end{figure}
Here, we quantified the extents of three types of changes introduced into the network structure during the CNS-based graph transformation process at different levels of noise.
A particularly relevant change to be examined here is the dropping of noisy edges by the different CNS measures during graph transformation. For this, we
recorded the percentage of noisy edges that were dropped in the transformed network generated using each of the measures at every noise level, and the results of this analysis are plotted in Figure
\ref{fig:removed_vs_noise}. Interestingly, although all the measures are able to eliminate a non-trivial fraction of the noisy edges, $HC.cont$ leads to the elimination of the highest fraction of
noisy edges, followed by $TOM.cont$ and $FS.cont$. However, it is important to examine this change in combination with other changes as well in order to obtain a more
comprehensive view of the CNS-based transformation procedure. Two other such important changes are the pruning of some of the non-noisy edges in the original network and the addition of new edges
(those not in the corresponding noisy network) into the transformed network. To study the extents of these changes, we collected the percentages of the number of non-noisy edges in the original
and the total number of edges in the transformed networks represented by these two types of edges to respectively. The variations
of these percentages at different levels of noise are shown in Figures~\ref{fig:valid_edges_dropped} and~\ref{fig:fraction_edges_added} respectively. A comparison of Figures~\ref{fig:removed_vs_noise} and
\ref{fig:valid_edges_dropped} shows that a smaller percentage of non-noisy edges are dropped by all the CNS measures as compared to the percentage of noisy edges dropped. For instance, $HC.cont$ drops
$~35-40\%$ of the noisy edges, as compared to $~28-30\%$ of the non-noisy edges, and the trend is similar for other measures also\footnote{In reality, not all the edges in the original network are 
non-noisy due to the inherent noise in the data (one of the motivations of this work). Thus, if a comparison is carried out using a set of perfectly non-noisy edges, this difference will
be larger.} This indicates that the CNS measures are indeed effective at differentiating between spurious and valid interactions on the basis of the common neighborhood information. Finally, it can
be observed from Figure~\ref{fig:fraction_edges_added} that all the CNS measures introduce a certain percentage of new edges into the transformed network to replace the noisy and non-noisy
edges dropped. For instance, about $30\%$ of the edges in the $HC.cont$-transformed network are new ones at all the noise levels tested here. Thus, the results shown in Figures
\ref{fig:removed_vs_noise}-\ref{fig:fraction_edges_added} show that several major changes are made to the original network during the graph transformation process by all the CNS measures. It
is particularly interesting to note that the ordering of these measures in terms of the extents to which they introduce these changes, namely $HC.cont$, $TOM.cont$ and $FS.cont$, is the same as
that for the function prediction results presented in Section 3.2 of the main text.

\subsection*{Global enhancement of functional coherence}

\begin{table}[t!]
\centering
{\footnotesize \begin{tabular}{|c|c|c|c|c|c|c|c|}\hline
\multirow{2}{*}{{\bf Network}} & \multicolumn{2}{|c|}{{\bf Common edges}} & \multicolumn{2}{|c|}{{\bf Dropped edges}} & \multicolumn{2}{|c|}{{\bf Added edges}} & \multirow{2}{*}{{\bf Overall}}\\
\cline{2-7}
& {\bf Fraction} & {\bf Avg. \#} & {\bf Fraction} & {\bf Avg. \#} & {\bf Fraction} & {\bf Avg. \#} & \\
& & {\bf functions shared} & & {\bf functions shared} & & {\bf functions shared} & \\\hline
$FS.cont$ & $0.9528$ & $0.9165$ & $0.0472$ & $0.2569$ & $0.0472$ & $1.469$ & $0.9421$\\\hline
$TOM.cont$ & $0.8839$ & $0.9461$ & $0.1161$ & $0.424$ & $0.1161$ & $1.235$ & $0.9797$\\\hline
$HC.cont$ & $0.7086$ & $1.0673$ & $0.2914$ & $0.4428$ & $0.2914$ & $0.8158$ & $0.9940$\\\hline
\end{tabular}}
\caption{Fraction and function relevance of different types of edges in the transformed networks.}
\label{table:functional_relevance}
\end{table}
The goal of this part of our study was the examine how the changes introduced into the network structure by the CNS-based graph transformation process influences the functional coherence of the
resultant transformed networks. For this, we categorized the different edges into three categories, namely
the edges common to the original and transformed networks (Common), those dropped from the original network during the transformation (Dropped) and those added to the transformed network to keep its size
(approximately) the same as the original network (Added). For each of these types of edges, we computed the average of the number of functions shared by the proteins connected
by the edges of that type\footnote{The trends reported here are consistent if we used at least one shared function as a measure of functional coherence.}. The
results of this analysis, along with the fractions of the transformed network represented by each type of edge, are presented in Table~\ref{table:functional_relevance}, and several trends can be
observed from them. First, although $HC.cont$ retains the smallest fraction of the original network, this subnetwork is also the most functionally coherent, followed by the $TOM.cont$ and $FS.cont$
retained subnetworks. This order is reversed for the cases of dropped and added edges, where $FS.cont$ drops the least functionally coherent edges and adds the most functionally coherent ones, followed
by $TOM.cont$ and $HC.cont$. Thus, there is substantial variation both in the fractions as well as the functional coherence of the sets of edges of these three types produced by the CNS measures
considered, which leads to the natural question of how these factors combine to determine the functional coherence of the final transformed networks? The answer to this question is provided by the
last column of Table~\ref{table:functional_relevance}, which shows the average number of functions shared by all pairs of connected proteins in the transformed networks. It is encouraging to note
that this measure of functional coherence is significantly higher for all the transformed networks when compared to the score of the original network ($0.8854$). More specifically, the $HC.cont$
transformed network has the highest functional coherence, followed by the $TOM.cont$ and $FS.cont$ networks, and these results match the order of performance of these measures in function
prediction (Tables $2$ and $3$ in the main text). This shows that the ability of $HC.cont$ to preserve the most functionally coherent part of the original network, and replace the
dropped edges with new edges that are reasonably functionally coherent leads to it producing the most functionally coherent transformed network. On the other hand, although $FS.cont$ adds more functionally
coherent edges, the fraction of these edges in the transformed network is very small, and thus the transformed network is not as functionally coherent. The performance of $TOM.cont$ is intermediate
from this point of view.

\end{document}